\documentclass[cmbright,fleqn,referee]{envauth}
\pdfoutput=1

\def\bSig\mathbf{\Sigma}

\received{00 Month 2012}
\revised{00 Month 2012}
\accepted{00 Month 2012}

\usepackage{natbib,upgreek}
\usepackage{mathrsfs}
\usepackage{tcolorbox}
\usepackage{bm}
\usepackage{bbm}
\usepackage{mathtools}

\runninghead{M. L. ~Bartley,  E. M. ~Hanks, and D. P. ~Hughes}{A Bayesian Penalized Hidden Markov Model}

\begin{document}

\title{A Bayesian Penalized Hidden Markov Model for Ant Interactions}

\author{M. L. Bartley\affil{a}\corrauth,  E. M. Hanks\affil{a}, and D. P. Hughes\affil{b}}

\corraddr{M. L. Bartley, Department of Statistcs, The Pennsylvania State University, University Park, PA, U.S.A. E-mail: bartley@psu.edu}

\address{\affilnum{a}Department of Statistics, The Pennsylvania State University, University Park, PA, U.S.A.\\
\affilnum{b}Department of Entomology, The Pennsylvania State University, University Park, PA, U.S.A.
}

\begin{abstract}
Interactions between social animals provide insights into the exchange and flow of nutrients, disease, and social contacts. We consider a chamber level analysis of trophallaxis interactions between carpenter ants (\textit{Camponotus pennsylvanicus}) over 4 hours of second-by-second observations. The data show clear switches between fast and slow modes of trophallaxis. However, fitting a standard hidden Markov model (HMM) results in an estimated hidden state process that is overfit to this high resolution data, as the state process fluctuates an order of magnitude more quickly than is biologically reasonable.  We propose a novel approach for penalized estimation of HMMs through a Bayesian ridge prior on the state transition rates while also incorporating biologically motivated covariates. This penalty induces smoothing, limiting the rate of state switching that combines with appropriate covariates within the colony to ensure more biologically feasible results. We develop a Markov chain Monte Carlo algorithm to perform Bayesian inference based on discretized observations of the contact network. 
\end{abstract}

\maketitle

\section{Introduction}
\label{s:intro}


Penalized estimation has long been a valuable tool in the development of models that provide biologically reasonable and interpretable results.  Ridge and LASSO penalties provide a means for variable selection and regulation in a linear model framework.  However, penalized estimation in the context of stochastic processes has yet to be fully explored. When exploring second by second ant behavior data we observed evidence of switching between periods of high and low rates of feeding interactions. When a standard Poisson hidden Markov model was applied to these data the resulting latent state predictions were overfit, or switching much faster than biologically reasonable.  We propose an approach to penalize stochastic processes within a hidden Markov model framework and apply this approach to high resolution behavioral interaction data. Our work provides a way to combat overfitting common in high resolution data through  Ridge or LASSO-like penalizing priors within a Bayesian context.

\subsection{Ant System}
Ants provide an ideal system for which to study social organisms in a controlled environment. Ants can be used as a proxy for learning about how humans in a social community deal with spread of disease, space utilization, and environmental variation (\cite{Fewell2013}). While we cannot conduct ethical controlled experiments on human populations, ant colonies are easy to maintain and manipulate in a laboratory setting. Researchers can control the space (size and organization), infections, nutrients, etc all while continuously monitoring behavioral interactions within the colony. By understanding how ants behave in these systems we may draw conclusions about our own interactions and abilities to mitigate spread of diseases. 

\begin{figure}
 \centerline{\includegraphics[width=6in]{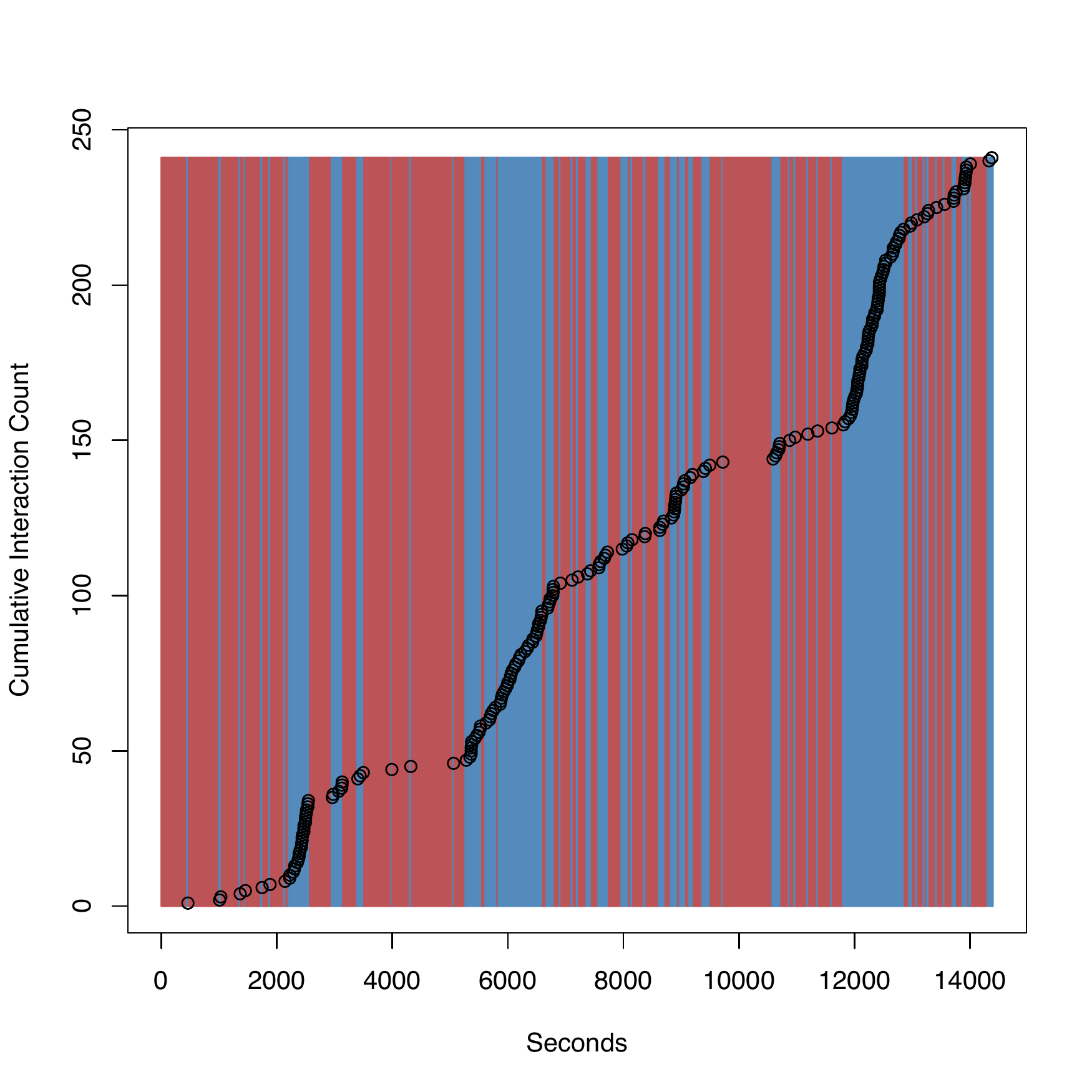}}
\caption{While periods of relatively low and high rates of feeding interactions are clear, the stochastic process of chamber-level feeding states is switching between the two much faster than biologically feasible. Red background denotes low state while blue background denotes high state of feeding exchanges. }
\label{f:simplestates}
\end{figure}

Ants engage in an oral exchange of nutrients called trophallaxis. Much like a hand shake or embrace in our own communities, these interactions in ants are a primary opportunity for disease transmission (\cite{Naug2002}) and for sharing nutrients through the colony. On both the colony and functional group level, ants have been shown to organize themselves, both spatially and temporally, in a way that defends against the spread of disease \citep{Quevillon2015}. It is imperative to be able to model interactions first under a healthy system to provide a basis for comparison when subsequent work manipulates the colonies through experimental infections. The Hughes Lab at Penn State maintains several wild-collected colonies of the black carpenter ant, \textit{Camponotus pennsylvanicus}, continuously monitored inside specially designed wooden chambers under complete darkness. Videos were examined frame by frame to identify and record each feeding event, the individual ants involved, and the start and end timing. In this study, we use four hours of these continuously monitored feeding interactions for one colony, and consider only interactions observed in the nest chamber in which the queen ant resides. Figure \ref{f:simplestates} shows the cumulative number of trophallaxis interactions observed in the four hours (14,400 seconds) of observations within the single chamber.

	An initial examination of the feeding interactions over time (see Figure \ref{f:simplestates}) suggests there may be periodic pulses in the rates of events. These pulses may reflect some latent underlying chamber-level behavioral states that are driving the different trophallaxis rates over time. For example, foraging ants re-entering the nest may spark a string of trophallaxis events as they seek to distribute nutrients through the colony. Previous research has also shown that ant colonies have been found to exhibit collective activity cycles (\cite{Richardson2017}). However, although we may observe evidence of these cycles, the mechanisms or causes of switching remain uncertain. The combination of high-resolution observations of each ant feeding exchange with these proposed unobserved chamber-level behavioral states lends itself nicely to exploration via a hidden Markov model. Hidden Markov models are often used to model ecological systems because they allow for behaviors to be correlated over time in a way that accounts for shifts in an underlying state process. HMMs have been used recently  to study animal movement behavior (\cite{Langrock2012,McKellar2015, Patterson2017,Towner2016, VandeKerk2015}), general animal behavior (\cite{DeRuiter2016,Langrock2014,Schliehe-Diecks2012}), as well as other applications within population ecology (\cite{Borchers2013,Gimenez2014, Johnson2016, Leos-Barajas2017a}). 
    
    When fitting a two-state Poisson HMM to the ant feeding interaction observations (described below in Section \ref{s:standard}), the estimated state switching rates were much too fast to be biologically reasonable (Figure \ref{f:simplestates}) (switching approximately 0.8 times per minute). In fact, the best estimated latent chamber level state process often switches between states right before and after individual interaction events resulting in fast estimated switching rates. This motivates the need to develop penalized approaches to fitting stochastic processes, like HMMs, to combat the apparent overfitting of the data. High resolution time series are more and more common, so this problem will be important in many future scientific applications. 

In this paper, we propose a novel penalized hidden Markov model that penalizes rapid state switching through  Bayesian ridge- or LASSO-like priors on continuous-time Markov chain transition rates. We show that this smooths the rate of estimated behavioral state switching and results in better 1-step ahead prediction than the non-penalized HMM. We demonstrate the utility of this penalized approach to stochastic process modeling on the ant trophallaxis data. We then extend this approach to model the effect of biological covariates, such as the arrival of forager ants into the colony or chamber, on transition rates between HMM states in order to better understand the cause of these behavioral switches. 

\section{Standard Hidden Markov Model}
\subsection{Model formulation}
\label{s:standard}

To start, we consider a simple, baseline model for the start times of ant trophallaxis interactions. It should be noted that while the duration of each feeding event was also recorded, we focus on modeling just the starting times of events, as this will provide insights into when and how often ant colonies exhibit pulses of interactions. Here we will outline the model for the case of $n=2$ latent states, and then cover how to extend to any $n$-state generalization of the model. We expect, based on prior observations, that the ants in the chamber of interest switch between states characterized by high and low rates of food-sharing events between ants. 
For any Markov-switching model the latent state process $X_t$ will only depend on the state ($X_{t - 1}$) the ants within the chamber were in at time $t - 1$. The state transition probabilities for this process are denoted by $p_{ij} = Pr(X_{t+1} = j | X_t = i)$ for $i, j = 1, \dots, n$. These probabilities are summarized in the transition probability matrix, $\mathbf{P}$, with $p_{ij}$ being the entry in row $i$ and column $j$. We adopt a Bayesian approach and specify Dirichlet priors for each row of $\mathbf{P}$, denoted by $\mathbf{p'}_\ell$ for each $\ell$ in $1, \dots, n$:  
\begin{equation}
\mathbf{p'}_\ell \sim \text{Dirichlet}(\bm{\theta}_\ell)
\end{equation}
where $\bm{\theta}$ is a real numbered matrix of values that acts as weights of the prior probabilities and $\bm{\theta}_\ell$ is a row vector of $\bm{\theta}$ containing the prior probabilities for state $\ell$. 

In the application to the ant colony and the chamber level interactions where we consider models with 2 unobserved states, these states correspond to relatively high and low rates of trophallaxis interactions. As we have very high resolution data (observations of $N_t$ every second for four hours), we may expect that the transition probabilities of staying in each state are relatively high, while transition probabilities of changing states are relatively low. That is, we would expect the chamber-level behaviors to persist for some time before switching states. 

As the number of feeding interactions ($N_t, t = 1, \dots, T$) are discretely valued and non-negative, we consider Poisson distributions for the state-dependent process, denoted by $N_t$. Given the chamber is in behavioral state $X_t$ at time $t$, the observed number of trophallaxis events initiated at time t is
\begin{equation}
N_t | X_t \sim \text{Pois}(\lambda_{X_t}).
\label{e:NgivenX}
\end{equation}
Note that $\lambda_{X_t}, X_t \in \{L, H\}$ depends on the behavioral state that the ants are in at time $t$. To maintain identifiability and prevent label switching, we will model the rate of feeding interactions as a baseline rate, $\lambda_L$, and increase in rate, $\tilde{\lambda}_H$ such that
\begin{equation}
\lambda_{X_t} = \lambda_L + \tilde{\lambda}_H \mathbbm{1}_{\{X_t = H\}}.
\label{e:lambdasplit}
\end{equation}
We assume a prior such that both rate parameters $\{\lambda_L, \tilde{\lambda}_H\}$ are distributed as gamma random variables with separate hyperparameters such that
\begin{equation}
\lambda_L \sim \text{Gamma}(a, b)
\end{equation}
\begin{equation}
\tilde{\lambda}_H \sim \text{Gamma}(c, d).
\end{equation}

Given a prior $X_0 \sim \text{Multinomial}(1, \bm{\pi})$, with $\bm{\pi} = (0.5, 0.5)$ on a starting chamber-level behavioral state, $X_0$, the posterior distribution is
\begin{equation}
[\{X_t\}, \{\lambda_L, \tilde{\lambda}_H\}, \mathbf{P} | \{N_t\} ] \propto \prod_{t = 1}^T \left( [N_t | X_t, \lambda_L, \tilde{\lambda}_H] [X_t | X_{t-1}, \mathbf{P}] \right) [X_0][\lambda_L] [\tilde{\lambda}_H] \prod^n_{\ell = 1}[\mathbf{p}_\ell].
\end{equation}


Following previous work, we consider a data augmentation (\cite{Tanner1987, VanDyk2001}) scheme, which relies on the fact that the sum of independent Poisson random variables is also Poisson distributed. Let the number of events at time t be
\begin{equation}
N_t = N_{Lt} + \tilde{N}_{Ht} \mathbb{I}_{\{X_t = H\}}.
\end{equation}
Where $N_{Lt} \sim \text{Pois}(\lambda_L)$ and $N_{Ht} \sim \text{Pois}(\tilde{\lambda}_H)$ are independent Poisson random variables.
Note that this is equivalent to \eqref{e:NgivenX} - \eqref{e:lambdasplit}. The subset $N_{Lt}$ becomes a baseline number of interactions starting at time $t$, and $N_{Ht}$ is the increase in interactions starting while the ant chamber is in behavioral state, $X_t = H$. Under a Bayesian approach for inference, the full-conditional distributions for $N_{Ht}$ and $N_{Lt}$ are available in closed form. When $X_t$ is in the low state, we know that $N_t = N_{L_t}$, and $N_{H_t} = 0$. When $X_t$ is in the high state, the full-conditional distribution of $N_{L_t}$ and $N_{H_t}$ is
\begin{equation}
\left. \begin{pmatrix}
    	N_{Lt}\\
        N_{Ht}
    	\end{pmatrix} \right| N_t, \lambda_L, \tilde{\lambda}_H  \sim 
\text{Multinom}\left(N_t,  \left(\frac{\lambda_L}{\lambda_L + \tilde{\lambda}_H}, \frac{\tilde{\lambda}_H}{\lambda_L + \tilde{\lambda}_H}\right)\right).
\label{e:splitNdist}
\end{equation}
With this data augmentation approach, the posterior distribution is now
\begin{equation}
\begin{aligned}
&[\{\mathbf{X}\}, \{\lambda_L, \tilde{\lambda}_H\}, \mathbf{P}, \{N_{Lt}, N_{Ht}\} | N_t  ] \propto \\
&\prod_{t = 1}^T \left( [N_t | N_{Lt}, N_{Ht}] [N_{Lt} | X_t, \lambda_L] [N_{Ht} | X_t = H, \lambda_L, \tilde{\lambda}_H][X_t | X_{t-1}, \mathbf{P}] \right) [X_0][\lambda_L] [\tilde{\lambda}_H] \prod^n_{\ell = 1}[\mathbf{p}_\ell]
\end{aligned}
\end{equation}
Extension of this model to the $n-$state case is straightforward. We apply similar data augmentation to split the observed dataset into $n$ incremental subsets, each with a corresponding incremental rate parameter $\tilde{\lambda}_k$. Again we would have a baseline $N_{1t}$ with subsequent subsets, $N_{2t}, \dots, N_{nt}$, with $N_t = \sum^{X_t}_{k = 1} N_{kt}$. Each subset of the observed data would also have its own corresponding $\lambda$ parameter such that
\begin{equation}
\lambda_t = \lambda_1 + \tilde{\lambda}_{2}\mathbb{I}_{\{X_t \neq 1\}} + \dots + \tilde{\lambda}_n \mathbb{I}_{\{X_t \neq 1:(n-1) \}}
\end{equation}
\begin{equation}
N_t \sim \text{Pois}(\lambda_{t})
\end{equation}
The full-conditional for ($N_{1t}, \dots, N_{nT}$) is, similar to \eqref{e:splitNdist}, multinomial with probabilities proportional to $\tilde{\lambda}_i$.
\begin{equation}
\left. \begin{pmatrix}
    	N_{1t}\\
        \vdots\\
        N_{nt}
    	\end{pmatrix} \right| N_t, \lambda_L, \tilde{\lambda}_H  \sim 
\text{Multinom}\left(N_t,  \left(\frac{\lambda_1}{\sum_{i=1}^{n}\lambda_{it}},
\dots,
\frac{\tilde{\lambda}_n}{\sum_{i=1}^{n}\lambda_{it}}\right)\right).
\end{equation}
\subsection{Model fitting}
To make inference on the model parameters described above $( \mathbf{\lambda}, \mathbf{P})$ and the latent path ($\mathbf{X}_{1:T}$), we constructed an MCMC algorithm to sample from the joint posterior distribution of all parameters. Conjugate updates were available for all parameters and so we were able to perform Gibbs updates for all parameter estimation. Hyperparameters for the above priors were chosen to be
 \begin{equation}
 a = 1,\quad
 b = 1,\quad
 c = 1, \quad
 d = 1, \quad
 \begin{pmatrix}
 \bm{\theta_L'}\\
 \bm{\theta_H'}
 \end{pmatrix} = \begin{pmatrix}
 120000, 1 \\
 1, 120000 
 \end{pmatrix}
 \end{equation} 
 To initialize the MCMC algorithm we first chose starting parameter values
 \begin{equation}
 \bm{P} = \begin{pmatrix}
 0.997 & 0.003 \\
 0.003 & 0.997
 \end{pmatrix}\qquad \bm{\lambda} = \{0.007, 0.05\}
 \end{equation}
 and initialized $\mathbf{X}$ for each time point. The algorithm was run for 50,000 iterations. 
 Chains were confirmed to have reached convergence through visual inspection. For the carpenter ant dataset, in the case of $n = 2$ states, this process in R (\cite{RCoreTeam2016}) takes 7 hours to run on a single core of a 2.76 Hz Intel Xeon Processor. 
\subsection{Results}
The resulting latent state estimates are presented in Figure~\ref{f:simplestates}. The posterior mean of $\lambda_L$, which represents the feeding event rate while the latent state is low, was $\hat{\lambda}_L$ = 0.00071 with a 95\% equal-tailed credible interval of (0.00007, 0.00192). The posterior mean for $\lambda_H$, the event rate when the latent state is high, was $\hat{\lambda}_H$ = 0.03838 with a 95\% equal-tailed credible interval of (0.03241, 0.04468). The estimate for $\lambda_L$ corresponds to a low chamber-level rate of feeding interactions where we would expect 0.04 trophallaxis interactions to start per minute. $\lambda_H$ corresponds to a high chamber-level rate of feeding events where we would expect to observed 2.3 interactions starting per minute. The transition probability matrix was estimated as 
\begin{equation}
\hat{\mathbf{P}} = \begin{pmatrix}
0.9857 & 0.0145\\
0.0145 & 0.9857 
\end{pmatrix}
\end{equation}
while the corresponding stationary distribution is $\hat{\bm{\delta}} = (0.503, 0.497)$. 
We can see in Figure~\ref{f:simplestates} that while this model succeeds in identifying two separate chamber level behavior states, it is clearly switching between these states far quicker than is biologically reasonable. 
Instead of identifying longer periods of relatively low and high chamber-level interactions rates, the high temporal resolution of the observations result in an overfit stochastic process (the latent Markov chain $X_t$) that switches from low to high at nearly every observed trophallaxis event and then switches back from high to low until the next event. Motivated by this overfitting, in the next phase of model development we develop a  penalized stochastic process, which combats the overfitting observed in this analysis. 
\section{A Penalized Stochastic Process Model}
\label{s:penalized}
\subsection{Model Formulating} 
The data show clear switches between fast and slow modes of trophallaxis; however, fitting a standard hidden Markov model (HMM) as in Section \ref{s:standard} results in an estimated hidden state process that is overfit to this high resolution data, as the state process fluctuates an order of magnitude more quickly than is biologically reasonable. To counter this overfitting, we propose a novel approach for penalizing stochastic processes, in particular discrete-space Markov chains, through Bayesian ridge and LASSO priors on the transition rates between states. These regularization priors induce smoothing of the stochastic process, limiting the rate of state switching to ensure more biologically interpretable results and better predictive power. 

While we have so far considered discrete-time Markov chains (DTMCs), we will develop penalized stochastic process models based on continuous time Markov chains (CTMCs), as they allow a straightforward approach for penalizing the rate of transition between states, and can serve as the basis for a DTMC model. Rather than modeling the transition probabilities $\mathbf{P}$ directly as we did in Section \ref{s:standard}, we model these probabilities as a function of state switching rates denoted by $\bm{\gamma} = (\gamma_{LH}, \gamma_{HL})$. Here $\gamma_{ij}$ is the rate of the ants in the monitored chamber switching from state $i$ to state $j$. Because the resolution of the data provide a second-by-second account of the ant fee	ding interactions, we consider a time-discretization of a continuous time Markov chain such that state switching may only occur on these same second intervals (i.e. only a times $t = 1, 2, \dots, T$). We chose not to model exclusively within a continuous time specification as there is no need to model at a sub-second resolution (higher resolution provides no additional benefit or interpretability) and discretization would need to occur when covariates are incorporated into the model. The discrete-time transition probabilities are
\begin{equation}
\begin{aligned}
p_{ij} &=  \text{Pr}(
\text{Colony in specified chamber remains in state $i$ for 1 second before switching to state $j$})\\
    &= \text{Pr}(\text{wait time in state $i$ is 1 second}) * \text{Pr}(\text{Colony in specified chamber switches to state $j$ after 1 second})\\
\end{aligned}
\label{e:pij}
\end{equation}
The wait time for a CTMC while in state $i$ is
\begin{equation}
W_{i} \sim \text{Exp}(\gamma_{i \cdot}), \quad \gamma_{i \cdot} = \sum_{\ell \neq i} \gamma_{i\ell}
\end{equation}
and the probability of switching to state $j$ from state $i$, given that the process leaves state $i$ is
\begin{equation}
 \text{Pr}(i \rightarrow j) = \frac{\gamma_{ij}}{\gamma_{i \cdot }}.
\end{equation}
Thus in this 2-state case, $p_{ij}$ in  \eqref{e:pij} will be equal to $\frac{\gamma_{ij}} {\gamma_{ij}} * \gamma_{ij} e^{-\gamma_{i\cdot}} = \gamma_{ij}e^{-\gamma_{i\cdot}}$ as long as this value is less than 1.   
With this CTMC-motivated approach in mind, we model our transition probabilities in the 2-state model (with 1 second temporal discretization) to be
\begin{equation}
\mathbf{P} = \begin{pmatrix}
1- \gamma_{LH}\exp\{- \gamma_{LH} \} & \gamma_{LH}\exp\{- \gamma_{LH} \}\\
\gamma_{HL} \exp\{- \gamma_{HL}\} & 1 - \gamma_{HL} \exp\{- \gamma_{HL}\}
\end{pmatrix}.
\label{e:2statePTM}
\end{equation}
Thus $\gamma_{HL}$ and $\gamma_{LH}$ control the rates of switching between behavioral states. We propose a penalized stochastic process model by considering Bayesian ridge and LASSO priors on these rate parameters. Under a Bayesian ridge prior, we model each rate $\gamma_{ij}$ as iid half-normal random variables
\begin{equation}
\gamma_{ij} \stackrel{iid}{\sim} \text{H. Norm}(0, \tau) 
\end{equation}
with prior density
\begin{equation}
[\gamma_{ij}] \propto \text{exp}\left\{\frac{-\gamma_{ij}^2}{2 \tau}\right\} \mathbbm{1}_{\{\gamma_{ij} > 0\}}.
\end{equation}
While we apply the ridge prior above, another option is to apply a LASSO prior, where each rate is modeled as exponentially distributed
\begin{equation}
\gamma_{ij} \stackrel{iid}{\sim} \text{Exp}\left(\frac{1}{\tau}\right)
\end{equation}
In both cases, $\tau$ is a tuning parameter that penalizes the overall rate of state switching in the Markov chain $\{X_t\}$. The smaller we make $\tau$, the stronger the penalty in the regularization. This is evident by considering the full-conditional distribution of the CTMC rate parameters.
Under the half normal prior, 
\begin{align}
\left[ \gamma_{LH}, \gamma_{HL} |X_t \right]
 &\propto \left(\prod_t [X_t | \mathbf{P}(\bm{\gamma})]\right) [\bm{\gamma}]_{\text{H. Norm}} \\
    &= \left(\prod_{t = 1}^T [\mathbf{P}_{X_t, X_{t+1}}]\right)  \exp\left\{\frac{-1}{2\tau} (\gamma_{LH}^2 + \gamma_{HL}^2) \right\}
\end{align}
Thus, as $\tau$ decreases there is increasing weight placed on slower CTMC transition rates. This results in a smoother stochastic process in which state transition rates are slower in general the smaller $\tau$ becomes.
For inference using MCMC we need to simulate the sample path of the chamber-level trophallaxis state Markov chain $X_{1:T}$ from its full conditional distribution:
\begin{equation}
\begin{split}
[X_{1:T}|N_{L\{1:T\}}, N_{H\{1:T\}}, \lambda_{X_t}] = &[X_T|N_{1:T}, \bm{\lambda}] [X_{T-1} | X_T, N_{1:T}, \bm{\lambda}]\\
&[X_{T-2} | X_T, X_{T-1}, N_{1:T}, \bm{\lambda}] \times \cdots \times [X_1|X_{T:2}, N_{1:T}, \bm{\lambda}]
\end{split}
\end{equation}
Following \cite{Zucchini2009a} we draw values backwards from $X_T$ to $X_0$ and in doing so require the following probabilities
\begin{equation}
[X_t | N_{L\{1:t\}}, N_{H\{1:t\}}, \lambda_{X_t}] \propto \alpha_t(X_t)
\label{e:startFB}
\end{equation}
where, 
\begin{align}
\bm{\alpha_t} &= (\alpha_t(1), \dots, \alpha_t(n))\\
\alpha_t(i) &= [N_{i1} = n_{i1}, \dots, N_{it} = n_{it}, X_t = i].
\end{align}
That is to say that each $\bm{\alpha_t}$ is a row vector with number of elements equal to the number of latent states (here we are considering two states).  These values may be computed from the recursion
\begin{align}
\bm{\alpha_1} = \bm{\pi} \begin{pmatrix}
[N_1|\lambda_{L}, X_1 = L] & 0\\
0 & [N_t | \lambda_{H}, X_1 = H]
\end{pmatrix}  \\
\bm{\alpha_t = \alpha_{t - 1}}\mathbf{P}\begin{pmatrix}
[N_t|\lambda_{L}, X_t = L] & 0\\
0 & [N_t | \lambda_{H}, X_t = H]
\end{pmatrix} .
\end{align}
Here, as before, $\bm{\pi}$ = (0.5, 0.5) is the the initial distribution of the Markov Chain. We first draw $X_T$ before simulating the remaining states in the order of $T - 1$ to 1. 
\begin{align}
[X_T | N_{L\{1:T\}}, N_{H\{1:T\}}, \lambda_{X_T}] &\propto \alpha_T(X_T)\\
[X_t | N_{L\{1:T\}}, N_{H\{1:T\}}, X_{T:t+1}, \lambda_{X_t}] &\propto \alpha_t(X_t) [X_{t+1} | X_t, \lambda_{X_t}]
\label{e:endFB}
\end{align}
The posterior distribution of the above HMM with penalized priors on $\bm{\gamma}$ is now
\begin{equation}
\begin{aligned}
 &[\{X_t\}, \{\lambda_L, \tilde{\lambda}_H\}, \{\gamma_{LH}, \gamma_{HL}\}, \{N_{Lt}, N_{Ht}\} | N_{X_t} ] \propto \\
 &\prod_{t = 1}^T \left( [N_{Lt} | X_t, \lambda_L] [N_{Ht} | X_t = H,  \tilde{\lambda}_H] [X_t | X_{t-1}, \mathbf{P}(\bm{\gamma})] \right) [X_0] [\lambda_L] [\tilde{\lambda}_H] [\gamma_{LH}][\gamma_{HL}]
\end{aligned}
\end{equation}
Extension of this model to the n-state case requires a similar data augmentation adjustment as seen in Section \ref{s:standard}. We now consider an $n \times n$ matrix $\mathbf{P}$ (with elements $p_{ij}$) with numbered states $1:n$ (rather than with our previous low/high convention), 
\begin{equation}
\bm{P} = \begin{pmatrix}
1 - p_{1\cdot} & \frac{\gamma_{12}}{\gamma_{1 \cdot}} * \gamma_{1\cdot}\exp\{- \gamma_{1\cdot}\}  & \cdots &  \frac{\gamma_{1n}}{\gamma_{1 \cdot}} * \gamma_{1\cdot}\exp\{- \gamma_{1\cdot}\}  \\
 \frac{\gamma_{21}}{\gamma_{2 \cdot}} * \gamma_{2\cdot}\exp\{- \gamma_{2\cdot}\}  &  \ddots & \cdots & \frac{\gamma_{2n}}{\gamma_{2 \cdot}}  * \gamma_{2\cdot}\exp\{- \gamma_{2\cdot}\}  \\
\vdots & \vdots & \ddots & \vdots \\
 \frac{\gamma_{n1}}{\gamma_{n \cdot}}  * \gamma_{n\cdot}\exp\{- \gamma_{n\cdot}\}  & \cdots & \cdots & 1 - p_{n \cdot}
\end{pmatrix}
\label{e:nstatePTM}
\end{equation}
where $p_{i\cdot} = \sum_{\ell \neq i} p_{i\ell}$ represent the off-diagonal row sums.
\subsection{Ensuring a Valid Probability Transition Matrix}

The probability transition matrices in \eqref{e:2statePTM} and \eqref{e:nstatePTM} will be valid as long as all off diagonal elements in a row sum to less than 1. We may ensure this requirement is met by exploring the maximum value of the sum of off diagonal elements. The critical points are found using calculus.  
\begin{align}
0 &= \frac{d}{d\gamma} \gamma_{i\cdot} e^{-\gamma_{i\cdot}\Delta}\\
&=\gamma_{i\cdot}(-\Delta)e^{-\gamma_{i\cdot}\Delta} + e^{-\gamma_{i\cdot}\Delta}\\
&= (1 - \gamma_{i\cdot}\Delta)e^{-\gamma_{i\cdot}\Delta}.
\end{align}
Here $\Delta$ is the time in which the ants within the chamber remain in state $i$ before switching to state $j$. The critical point of $\gamma_{i\cdot} e^{-\gamma_{i\cdot}\Delta}$ occurs at $\gamma_{i\cdot} = \frac{1}{\Delta}$ and evaluating the second derivative at this point confirms that it is a local maximum. So if we want to ensure that our off diagonal elements sum to less than 1 in the n-state case we must ensure that
\begin{equation}
1 \geq \sum_{j \neq i}\left(\gamma_{ij}e^{-\gamma_{i\cdot}\Delta} \right) = \gamma_{i\cdot}e^{-\gamma_{i\cdot}\Delta}
\end{equation}
        Plugging in the maximum, $\gamma_{i\cdot} = \frac{1}{\Delta}$ yields $1 \geq \frac{1}{\Delta}e^{-1}$. This equality holds as long as $\Delta \geq e^{-1}$. In our analysis we have $\Delta = 1$ second and this condition is met for any set of rate parameters $\{\gamma_{ij}\}$. 
\subsection{Model fitting}
This penalized hidden Markov model for ant feeding events aims to improve biological interpretability and predictive power over the standard HMM, which we showed in Section \ref{s:standard} is prone to overfitting when applied to high-resolution data. To make inference on the model parameters $(\gamma_{LH}, \gamma_{HL}, \lambda_L, \tilde{\lambda}_H)$ and the latent state path $(\mathbf{X}_{1:T})$ we constructed a MCMC algorithm to sample from the posterior distribution above. The hyperparameters detailed above were chosen to be 
\begin{equation}
a = 1, \qquad
b = 1, \qquad
c = 1, \qquad
d = 1. \qquad
\end{equation}
With this algorithm, we first update the switching rates, $\bm{\gamma}$. We conduct a normal random walk Metropolis Hastings update centered on the log of $\bm{\gamma}$. This allows for Metropolis-Hastings updates of the probability transition rates, $\bm{\gamma}$. All proposal distributions were tuned adaptively using the log adaptive proposals of \cite{Shaby2010}. Subsequently, at each iteration of the MCMC algorithm, we sample new values for $X_{1:T}$ through the previously described forwards/backwards algorithm \eqref{e:startFB} - \eqref{e:endFB}, and also sample ($\lambda_L, \tilde{\lambda}_H$) with Gibbs updates similar to those described in \ref{s:standard}. 

While we apply our novel two-state, penalized hidden Markov model to trophallaxis data as described above,  we also considered 3-state models. Here, we are estimating values for three feeding interaction rates, \{$\lambda_L, \tilde{\lambda}_M, \tilde{\lambda}_H$\}, and six stochastic process state switching rates \{$\gamma_{LM}, \gamma_{LH}, \gamma_{ML}, \gamma_{MH}, \gamma_{HL}, \gamma_{HM}$\}. Here the hyperparameters for the Gamma distribution priors of the $\lambda$ variables were chosen to be 
\begin{equation}
a = 1, \qquad
b = 1, \qquad
c = 1, \qquad
d = 1, \qquad
e = 1, \qquad
f = 1. \qquad
\end{equation}
\subsection{Choosing The Tuning Parameter $\tau$}
\label{ss:compare}
\begin{figure}
 \centerline{\includegraphics[width=7in]{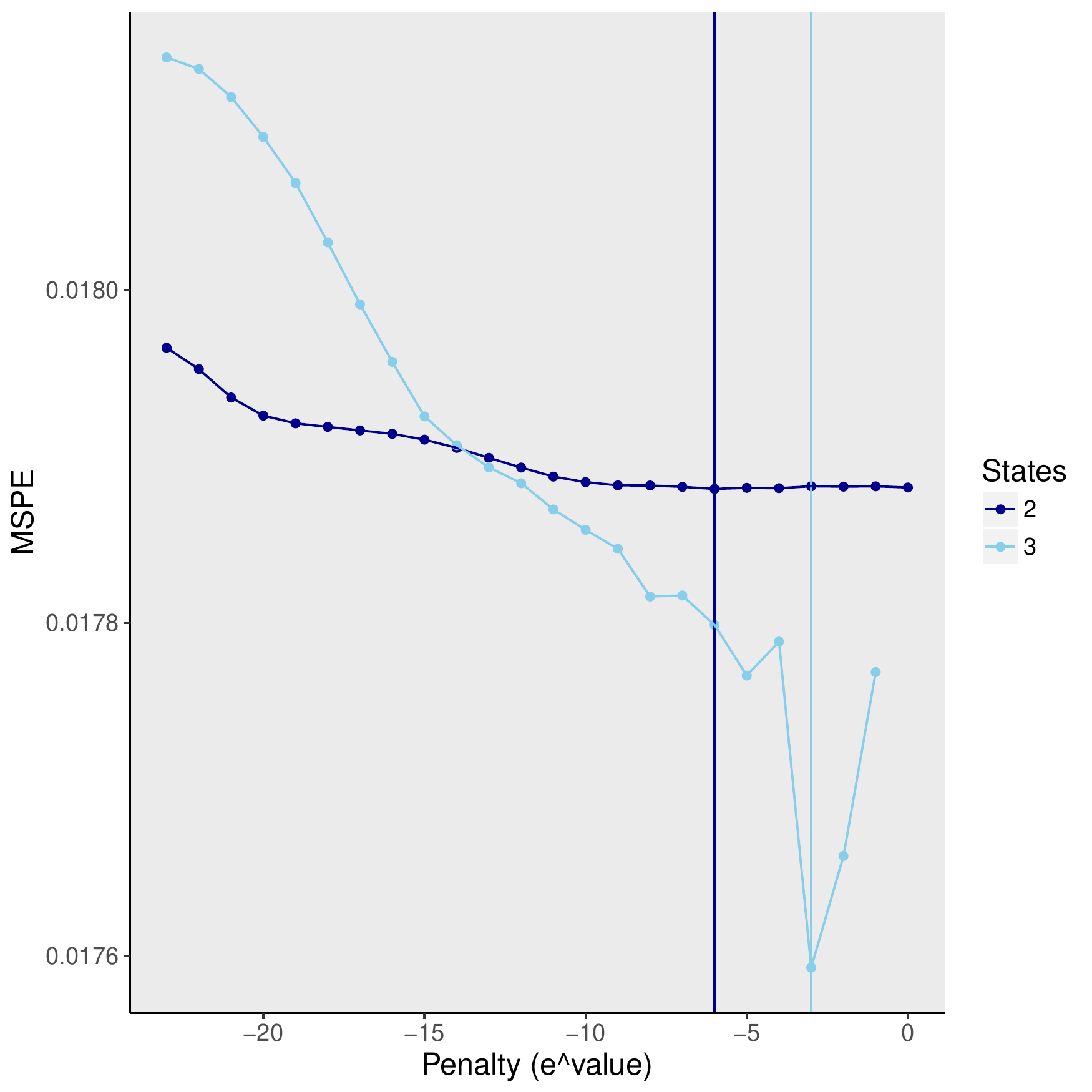}}
\caption{MSPE Comparison results }
\label{f:MSPEpen2&3}
\end{figure}

The choice of the prior variance of $\bm{\gamma}$ is an important one as it controls the penalization and smoothness of $X_t$. We explore the space of $\tau$ to ensure the best predictive model has been specified. Model comparison is accomplished through comparison of one-step ahead posterior predictive mean squared prediction error (MPSE). 
\begin{equation}
MSPE(\tau) = E\left[ \sum_t (\hat{N_t} - N_t) | \bm{N}\right]
\end{equation}
where
\begin{align}
\hat{N}_{t+1}&= E\left[ N_{t+1} | X_t \right]
\\
&= \sum_{k=1}^n \lambda_k P_{X_t, k}
\end{align}
We approximate MSPE($\tau$) using draws of \{$\lambda_k, \{X_t\}, \mathbf{P}$\} from the posterior.  Each penalized HMM was run to convergence for a range of potential $\tau$ values. The best predictive model was chosen to be the one with the minimized MSPE value. 
As illustrated in Figure \ref{f:MSPEpen2&3}, MSPE for the two state penalized HMM was minimized at $\tau = e^{-6}$ while it was minimized at $\tau = e^{-3}$ for the model with three states specified. 
Figure \ref{f:MSPEpen2&3} shows that the optimized three state model out performs the two state model when using estimated mean square prediction error as the means of model comparison. Selecting a number of biological states for hidden Markov models has been shown to be difficult (\cite{Pohle2017}) where additional states may capture some ignored data structure within the model. 
\subsection{Results}
\begin{figure}
 \centerline{\includegraphics[width=6in]{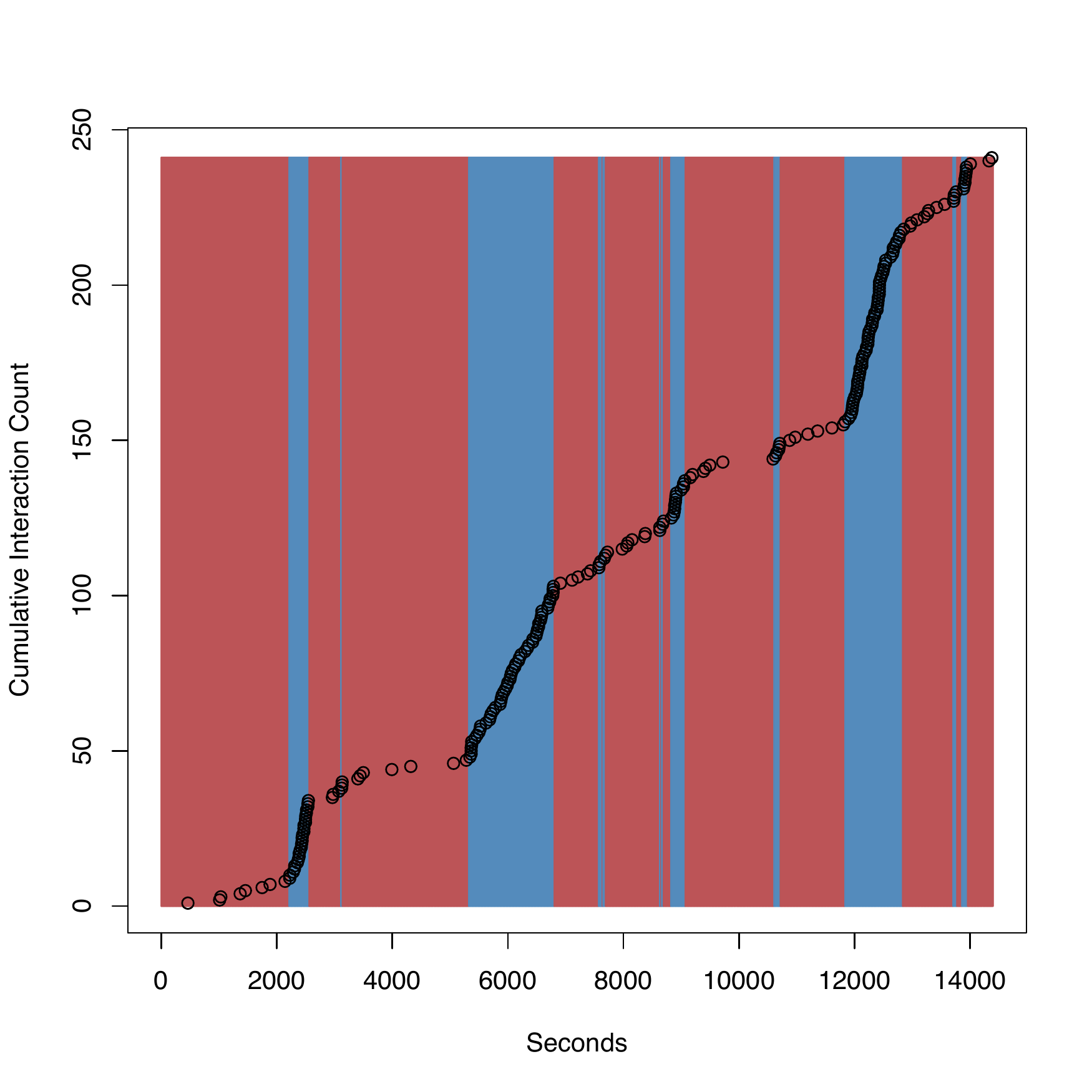}}
 \caption{Two state enalized HMM results. Again, red background denotes low state while blue background denotes high state of feeding exchanges.}
\label{f:penaltystates}
\end{figure}

The results that follow are the models with the optimized $\tau$ penalty parameter values as described in Section \ref{ss:compare}. The resulting posterior mean latent process estimates for the 2-state penalized HMM are presented in Figure \ref{f:penaltystates}. The posterior mean of $\lambda_L$, which represents the feeding event rate while the latent state is low, was $\hat{\lambda}_L = 0.0057$ with a credible interval of (0.00382, 0.00774). The posterior mean for $\lambda_H$, the event rate when the latent state is high, was $\hat{\lambda}_H = 0.0501$ with a credible interval of (0.04083, 0.06133). As in Section \ref{s:standard}, the estimate for $\lambda_L$ corresponds to a low chamber-level rate of feeding interactions where we would expect 0.34 trophallaxis interactions to start per minute. $\lambda_H$ corresponds to a high chamber-level rate of feeding events where we would expect to observed 3.01 interactions starting per minute. The posterior mean of $\gamma_{LH}$, which represents the chamber-level rate of state switching from Low to High, was $\hat{\gamma}_{LH} = 0.00142$  with a credible interval of (0.00050, 0.00289). The posterior mean for $\gamma_{HL}$, the state switching rate from High to Low, was $\hat{\gamma}_{HL} =  
0.00422 $ with a credible interval of (0.00156, 0.00874). From these $\bm{\gamma}$ estimates, we may calculate the posterior mean transition probability matrix 
\begin{equation}
\hat{\mathbf{P}} = \begin{pmatrix}
 0.9986 & 0.0014 \\
 0.0042 &   0.9958
\end{pmatrix}
\end{equation}
while the corresponding stationary distribution is $\hat{\bm{\delta}} = (0.75, 0.25)$. 
 We can see in Figure \ref{f:penaltystates} that  this model succeeds in limiting the rate of switching between high and low levels of ant chamber-level trophallaxis. This penalized model provides a more biologically reasonable rate of switching in addition to providing trophallaxis state predictions that align with a visual inspection of interactions over time. 
\begin{figure}
 \centerline{\includegraphics[width=6in]{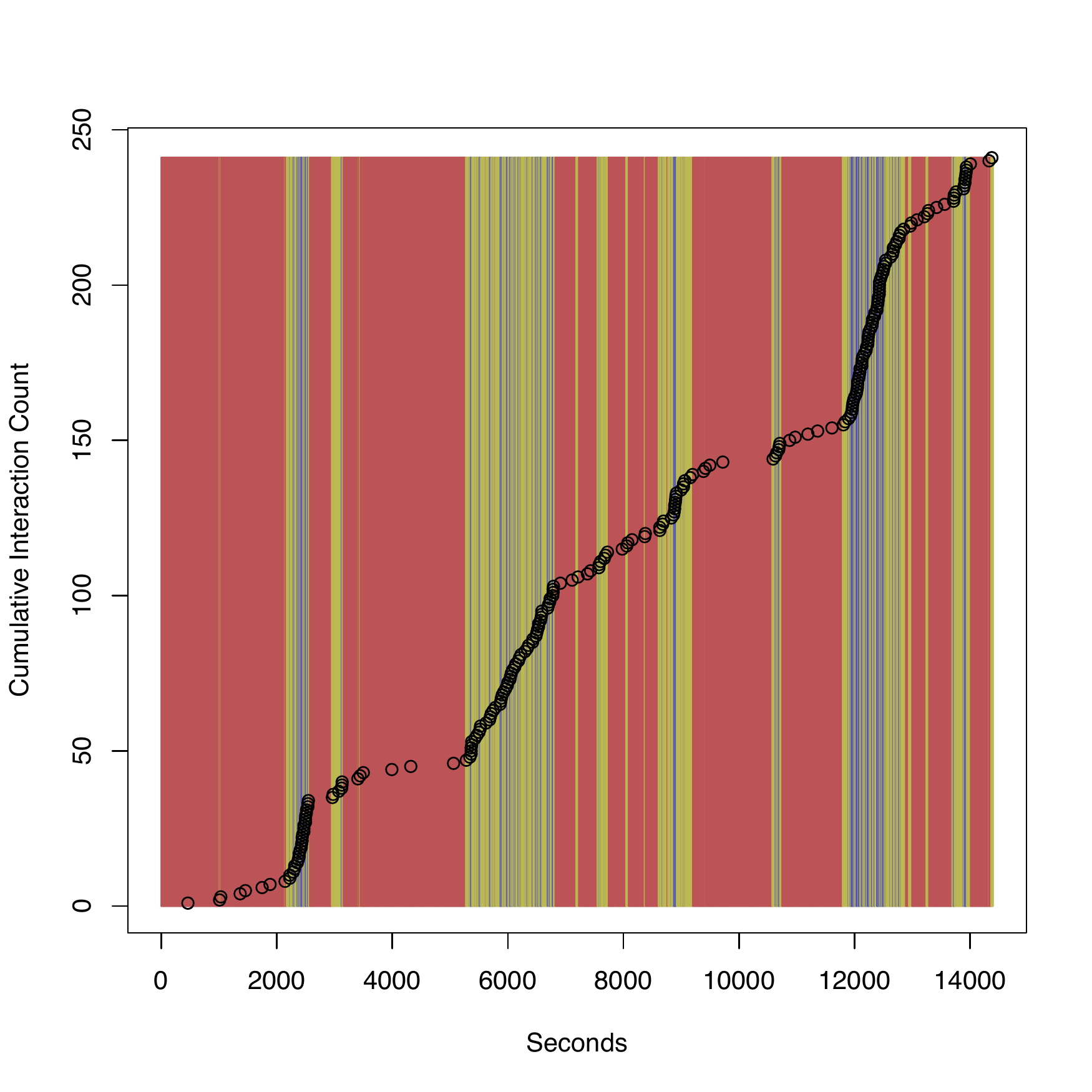}}
\caption{Three state penalized HMM results. Red background denotes low state while blue background denotes high state of feeding exchanges. Additionally, green now denotes the third, or medium, state of feeding exchanges.}
\label{f:pen3states}
\end{figure}
The chamber-level trophallaxis state estimates from the optimized 3-state penalty HMM model may be seen in Figure \ref{f:pen3states}. Interestingly, when the 3-state model was applied we found that the feeding interaction rates for the low and medium chamber-level state switching do not differ much ($\hat{\lambda_L} = 0.00351, \hat{\lambda_M} = 0.00404$)  while the feeding interaction rate for the high state is greater than as estimated in the two-state model ($\hat{\lambda_H} = 0.08663$). This results in effectively two states (low/medium \& high) but instead of the biologically reasonable switching as seen in Figure \ref{f:penaltystates}, the three state model exhibits switching around individual feeding interactions as seen in Figure \ref{f:simplestates} with the standard HMM approach. 
The estimated probability transition matrix is shown below \eqref{e:3stateP}. Once in the medium trophallaxis rate state, the colony is more likely to enter the high state when switching. Similarly, upon leaving the high state, the colony is more likely to re-enter the medium feeding rate state than the low interaction rate state..
\begin{equation}
\hat{\mathbf{P}} = \begin{pmatrix}
 0.9975 & 0.0010 & 0.0015 \\
 0.0063 & 0.7611 & 0.2327 \\
 0.0046 & 0.2243 & 0.7711
\end{pmatrix}
\label{e:3stateP}
\end{equation}
The corresponding stationary distribution is $\bm{\hat{\delta}} = (0.684, 0.153, 0.163)$.
\section{Penalized Model with Covariate(s)}
\label{s:covariate}
\subsection{Formulating the model}
We have shown that though penalizing the stochastic process we have limited the rate of state switching resulting in more biologically interpretable results and better predictive power. However we want to explore further to better explain what is causing these state switching in the ant colony. The availability of nutrients, the movement of forager ants through the colony, and the number of still hungry ants all may influence the rate of feeding events in the colony (at the chamber level). We extend our hidden Markov model to consider various biological covariates into our novel approach for penalizing stochastic processes. 
With this extension to include covariates within the model, we continue to model transition probabilities as a function of continuous-time state switching rates. However, our state switching rates are now a function of the covariate(s) and parameters. As the state switching rates must be non-negative, we consider a log transformation of a linear function of  covariates. Note that now our $\bm{\gamma}$ values (and by extension our transition probabilities) may vary over time, $t$, as we consider covariates that vary over time. Let $\bm{w}_t$ be a vector of covariates (not including an intercept). Then let 
\begin{equation}
\gamma_{ijt} = e^{\mu_{ij} + \bm{w}_t' \bm{\beta}_{ij}}.
\end{equation}
\begin{figure}
 \centerline{\includegraphics[width=5in]{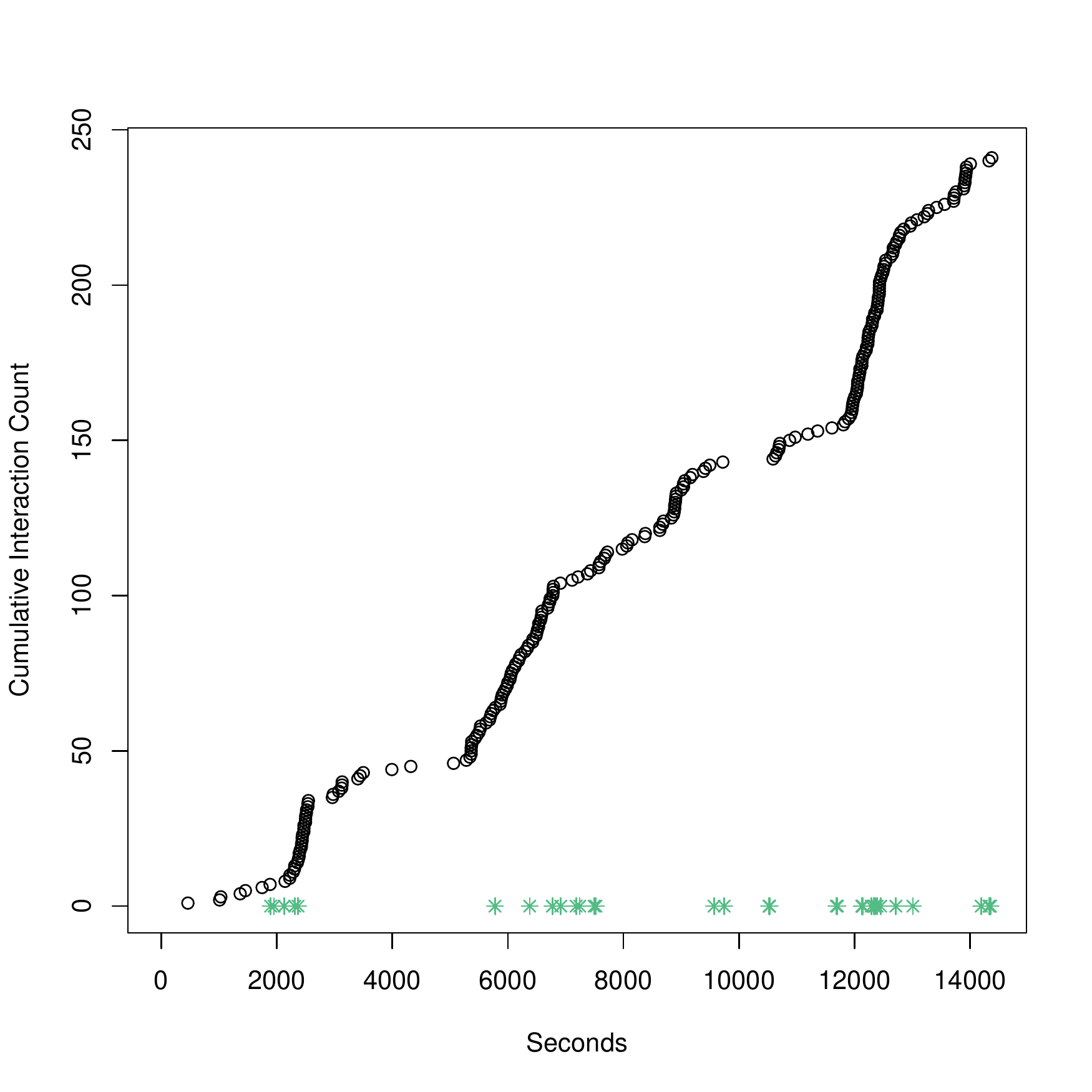}}
 \caption{Ant interaction data with cumulative feeding interactions over 4 hours of observation. Green asterisk symbols denote times at which an ant enters into the nest chamber.}
\label{f:pencov}
\end{figure}
For our ant system, we consider a single covariate, $w_t$, which is 1/(time since a foraging ant has entered the nest chamber). These entrance times may be seen in Figure \ref{f:pencov}. In the 2-state setting,
\begin{equation}
\begin{aligned}
\log (\gamma_{ijt}) &= \mu_{ij} + \beta_{ij} \left(\frac{1}{w_t^{\alpha} + 1}\right) \\
\gamma_{LHt} &= e^{\mu_{LH}} e^{\beta_{LH}\left(\frac{1}{w_t^{\alpha} + 1}\right)}\\
\gamma_{HLt} &= e^{\mu_{HL}} e^{\beta_{HL}\left(\frac{1}{w_t^{\alpha} + 1}\right)}\\
\end{aligned}
\end{equation}
With this updated definition of $\bm{\gamma}$ we have $e^{\mu_{ij}}$ in the same role of a baseline rate of state switching between states $i$ and $j$. While in Section \ref{s:penalized} our prior was placed on the entire rate of switching, here we propose a Bayesian ridge prior on the baseline rate, $e^{\mu_{ij}}$. 
\begin{equation}
 e^{\mu_{ij}} 
c(\beta_{ij})
\sim 
\text{H. Norm}(0, \tau)
\Rightarrow 
e^{\mu_{ij}} | \beta_{ij} \sim \text{H. Norm}\left(0, \frac{\tau}{c(\beta_{ij})^2}\right)
 \end{equation}
 where $c(\beta_{ij})$ is proportional to the expected number of transitions between chamber-level states.
\begin{equation}
  c(\beta_{ij}) =  \frac{1}{T}\sum_{t = 1}^T e^{\beta_{ij}\left(\frac{1}{w_t^{\alpha} + 1}\right)}
\end{equation}
Normalizing $\tau$ by dividing by $c(\beta_{ij})^2$ standardizes the penalization so that tau penalizes the overall expected rate of transitions. Here, $\tau$ still penalizes the rate of switching between high and low rates of trophallaxis in the colony as seen in Section \ref{s:penalized}. 

The remaining switching rate parameters $\{\beta_{ij}\}$ control the effect of $w_t$ on switching rates. When an ant enters into the chamber $w_t$ drops to zero resulting in a temporary increase in the rate of switching between states. The exponent parameter, $\alpha$ controls the rate of decay after this increase occurs. We assign both normal priors with hyperparmeters such that
\begin{equation}
\begin{aligned}
\beta_{ij} \sim \text{N}(1, 100)\\
\alpha \sim \text{N}(1, 10).
\end{aligned}
\end{equation}
\section{Model fitting}
This penalized hidden Markov model with biological covariates for ant feeding events aims to improve biological interpretability and predictive power over the basic and prediction-only models. To make inference on the model parameters $(e^{\mu_{LH}}, e^{\mu_{HL}}, \beta_{LH}, \beta_{HL}, \lambda_L, \tilde{\lambda}_H, \alpha)$ and the latent state path $(\mathbf{X}_{1:T})$ we constructed a MCMC algorithm to sample from the posterior distribution above. With this algorithm, we first update the switching rate parameters, \{$e^{\mu_{LH}}, e^{\mu_{HL}}, \beta_{LH}, \beta_{HL}, \alpha$\} jointly. We conduct a Normal random walk Metropolis Hastings update centered on \{$\mu_{LH}, \mu_{HL}, \beta_{LH}, \beta_{HL}, \alpha$\} using $\Sigma$ as the proposal variance. This allows for Metropolis-Hastings updates of the probability transition rates, $\bm{\gamma}$. All proposal distributions were tuned adaptively using again using Shaby and Wells' log adaptive proposals tuning. Subsequently, at each iteration of the MCMC algorithm, we sample new values for $X_{1:t}$ and ($\lambda_L, \tilde{\lambda}_H$) with the forwards/backwards algorithm and Gibbs updates, respectively, similar to those described in Section \ref{s:penalized}. 
\subsection{Results}
\begin{figure}
 \centerline{\includegraphics[width=6in]{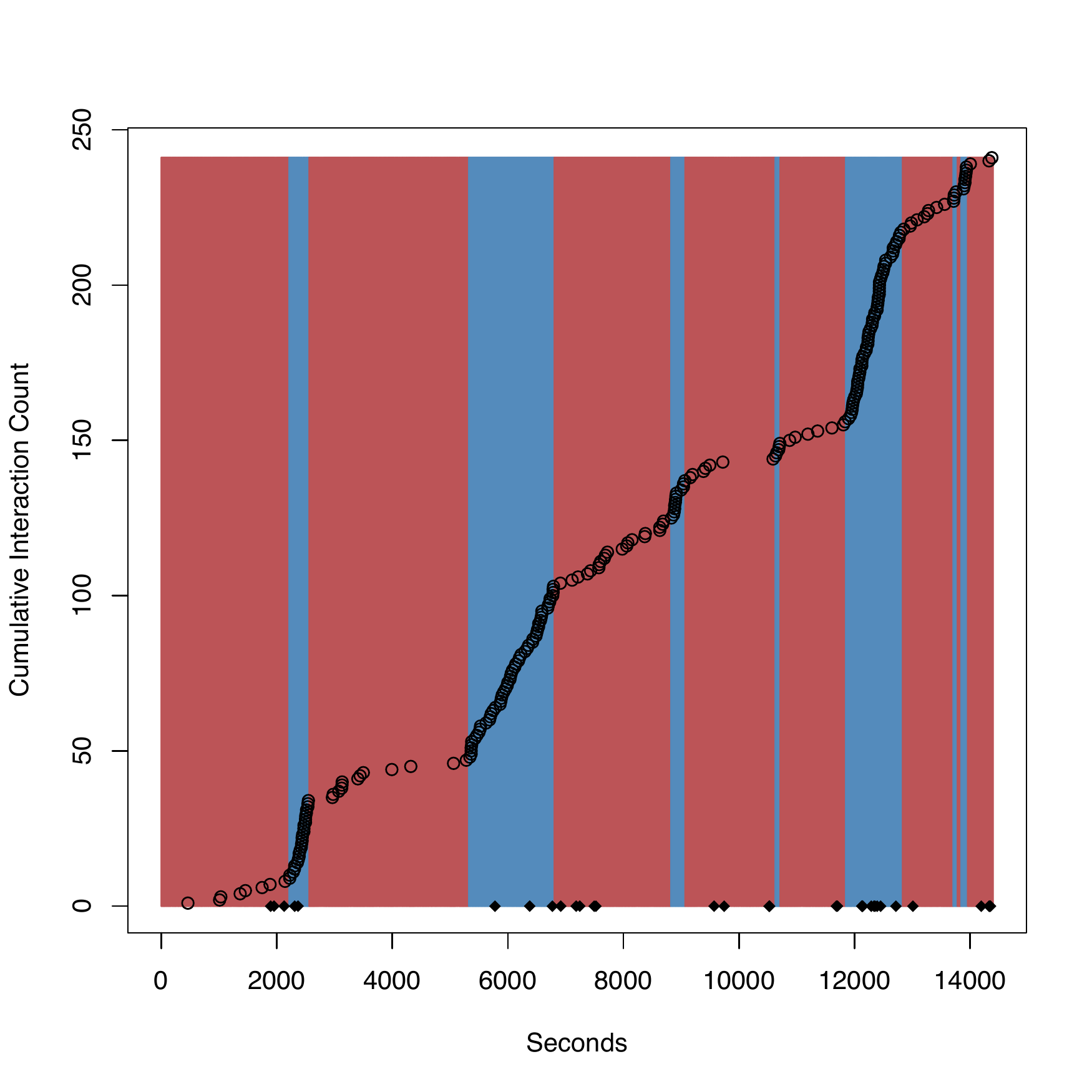}}
 \caption{Two state penalized HMM results with biological covariates - time since ant has entered the chamber. Again, red background denotes low state while blue background denotes high state of feeding exchanges. Black symbols denote entrance times.}
\label{f:covstates}
\end{figure}

The resulting latent process estimates are presented in Figure \ref{f:covstates}. The posterior mean of $\lambda_L$, which represents the feeding event rate while the latent state is low, was $\hat{\lambda}_L$ = 0.0062 with a credible interval of (0.00416, 0.00834). The posterior mean for $\lambda_H$, the event rate when the latent state is high, was $\hat{\lambda}_H$ = 0.0500 with a credible interval of (0.04113, 0.06036). As in \ref{s:penalized}, the first State corresponds to a low chamber-level rate of feeding interactions where we would expect 0.37 trophallaxis interactions to start per minute. State 2 corresponds to a high chamber-level rate of feeding events where we would expect to observed 3 interactions starting per minute. 

The posterior mean of $e^{\mu_{LH}}$, which represents the chamber-level rate of state switching from Low to High, was $\hat{e^{\mu_{LH}}} = 0.00087$  with a credible interval of (0.00005, 0.00210). The posterior mean of $\beta_{LH}$, which represents our covariate coefficient when switching from Low to High, was $\hat{\beta}_{LH}$ = -0.33285 with a credible interval of (-2.47636, 1.08674). The posterior mean for $e^{\mu_{HL}}$, the state switching rate from High to Low, was $\hat{e^\mu_{HL}} = $ 0.00470 with a credible interval of (0.00036,	0.01716). The posterior mean of $\beta_{HL}$, which represents our covariate coefficient when switching from High to low , was $\hat{\beta}_{HL}$ = -0.29110 with a credible interval of (-2.25673, 1.55876). The posterior mean of $\alpha$, which represents the rate of decay after an increase in $e^{\mu_{ij}}$, was $\hat{\alpha} = 0.0929$ with a 95\% credible interval of $(-2.50901, 1.73644)$. 

We can see in Figure \ref{f:covstates} that  this model succeeds in smoothing the overfitting of the stochastic process similarly to previous 2-state results. However we do not find evidence that the chosen covariate, ant entrance times into the chamber, are affecting the switching rates over time. The posterior distributions for both $\beta_{LH}$ and $\beta_{HL}$ contained zero. It is evident that this penalized model expanded to include biological covariates maintains the ability of penalizing the stochastic processes while testing biological hypotheses.
\section{Discussion}
\label{s:discuss}
We have shown that by penalizing stochastic processes through Bayesian ridge priors on the transition rates between trophallaxis rates states that we are able to counter overfitting common in high resolution data. By reducing our prior variance for the state switching rates, $\bm{\gamma}$, we increased the effective penalty and induced similar shrinkage on $\bm{\gamma}$ as in Bayesian ridge regression. We propose that similar results may be obtained through LASSO priors on the transition rates, but do not explore this alternative prior choice within our application to ant feeding interaction data. 

Other unexplored alternatives relevant this research that are beyond the scope of this paper include other biological covariates and different means of model comparison.  Colony or chamber level movement characteristics (avg. number/proportion of moving ants, etc) are available from the continuous observation of the colony and may provide further insight into the mechanisms/causes of this behavior switching. In addition, while the use of one-step ahead MSPE is computationally convenient, other model comparison methods may be considered such as a log-predictive score. While these are simple extensions to our proposed model, each would increase the computational requirements and possible increase the number of parameters to be estimated. Further development of this penalized HMM will provide researchers with a method of combating overfitting in high resolution data.      

Penalization is an important avenue of research as it has vast applications within both statistical and ecological fields. As animal interactions, behaviors, and movements are monitored at increasingly higher resolutions the novel approach outlined in this paper may be a useful tool. Current research with second-by-second monitoring of animals includes the work of \cite{Farine2016} with the study of the collective movement of a troop of wild olive baboons. Our application to ant trophallaxis behavior in the colony at the chamber level may easily be applied to bee colonies that also have exhibited bursty interaction patters (\cite{Gernat2017}), and other ecological data.


\section*{Acknowledgements}
Funding for this work is provided by NIH GM116927-01 and NSF EEID 1414296. 
\bibliographystyle{wb_env}
\bibliography{Mendeley}

\begin{thebibliography}{24}
\providecommand{\natexlab}[1]{#1}

\bibitem[{Borchers \textit{et~al.}(2013)Borchers, Zucchini, Heide-J{\o}rgensen,
  Ca{\~{n}}adas and Langrock}]{Borchers2013}
Borchers DL, Zucchini W, Heide-J{\o}rgensen MP, Ca{\~{n}}adas A, Langrock R,
  2013. {Using Hidden Markov Models to Deal with Availability Bias on Line
  Transect Surveys}. \textit{Biometrics} \textbf{69}(3): 703--713.

\bibitem[{DeRuiter \textit{et~al.}(2016)DeRuiter, Langrock, Skirbutas,
  Goldbogen, Chalambokidis, Friedlaender and Southall}]{DeRuiter2016}
DeRuiter SL, Langrock R, Skirbutas T, Goldbogen JA, Chalambokidis J,
  Friedlaender AS, Southall BL, 2016. {A multivariate mixed hidden Markov model
  to analyze blue whale diving behaviour during controlled sound exposures} :
  1--26.

\bibitem[{Farine \textit{et~al.}(2016)Farine, Strandburg-Peshkin, Berger-Wolf,
  Ziebart, Brugere, Li and Crofoot}]{Farine2016}
Farine DR, Strandburg-Peshkin A, Berger-Wolf T, Ziebart B, Brugere I, Li J,
  Crofoot MC, 2016. {Both nearest neighbours and long-term affiliates predict
  individual locations during collective movement in wild baboons}.
  \textit{Scientific Reports} .

\bibitem[{Fewell(2013)}]{Fewell2013}
Fewell JH, 2013. {References and Notes}. \textit{The Hunt for the Parathyroids}
  \textbf{301}(September): 132--143.

\bibitem[{Gernat \textit{et~al.}(2017)Gernat, Rao, Middendorf, Dankowicz,
  Goldenfeld, Robinson, Holme, Naug and Sokolowski}]{Gernat2017}
Gernat T, Rao VD, Middendorf M, Dankowicz H, Goldenfeld N, Robinson GE, Holme
  P, Naug D, Sokolowski MB, 2017. {Automated monitoring of behavior reveals
  bursty interaction patterns and rapid spreading dynamics in honeybee social
  networks} : 1--6.

\bibitem[{Gimenez \textit{et~al.}(2014)Gimenez, Blanc, Besnard, Pradel,
  Doherty, Marboutin and Choquet}]{Gimenez2014}
Gimenez O, Blanc L, Besnard A, Pradel R, Doherty PF, Marboutin E, Choquet R,
  2014. {Fitting occupancy models with E-SURGE: Hidden Markov modelling of
  presence-absence data}. \textit{Methods in Ecology and Evolution}
  \textbf{5}(6): 592--597.

\bibitem[{Johnson \textit{et~al.}(2016)Johnson, Laake, Melin and
  DeLong}]{Johnson2016}
Johnson DS, Laake JL, Melin SR, DeLong RL, 2016. {Multivariate state hidden
  Markov models for mark-recapture data}. \textit{Statistical Science}
  \textbf{31}(1958): 233--244.

\bibitem[{Langrock \textit{et~al.}(2012)Langrock, King, Matthiopoulos, Thomas,
  Fortin and Morales}]{Langrock2012}
Langrock R, King R, Matthiopoulos J, Thomas L, Fortin D, Morales JM, 2012.
  {Flexible and practical modeling of animal telemetry data: hidden Markov
  models and extensions.} \textit{Ecology} \textbf{93}(11): 2336--42.

\bibitem[{Langrock \textit{et~al.}(2014)Langrock, Marques, Baird and
  Thomas}]{Langrock2014}
Langrock R, Marques TA, Baird RW, Thomas L, 2014. {Modeling the Diving Behavior
  of Whales: A Latent-Variable Approach with Feedback and Semi-Markovian
  Components}. \textit{Journal of Agricultural, Biological, and Environmental
  Statistics} \textbf{19}(1): 82--100.

\bibitem[{Leos-Barajas \textit{et~al.}(2017)Leos-Barajas, Photopoulou,
  Langrock, Patterson, Watanabe, Murgatroyd and
  Papastamatiou}]{Leos-Barajas2017a}
Leos-Barajas V, Photopoulou T, Langrock R, Patterson TA, Watanabe YY,
  Murgatroyd M, Papastamatiou YP, 2017. {Analysis of animal accelerometer data
  using hidden Markov models}. \textit{Methods in Ecology and Evolution}
  \textbf{8}(2): 161--173.

\bibitem[{McKellar \textit{et~al.}(2015)McKellar, Langrock, Walters and
  Kesler}]{McKellar2015}
McKellar AE, Langrock R, Walters JR, Kesler DC, 2015. {Using mixed hidden
  Markov models to examine behavioral states in a cooperatively breeding bird}.
  \textit{Behavioral Ecology} \textbf{26}(1).

\bibitem[{Naug and Camazine(2002)}]{Naug2002}
Naug D, Camazine S, 2002. {The role of colony organization on pathogen
  transmission in social insects}. \textit{Journal of Theoretical Biology}
  \textbf{215}(4): 427--439.

\bibitem[{Patterson \textit{et~al.}(2017)Patterson, Parton, Langrock,
  Blackwell, Thomas and King}]{Patterson2017}
Patterson TA, Parton A, Langrock R, Blackwell PG, Thomas L, King R, 2017.
  {Statistical modelling of individual animal movement: an overview of key
  methods and a discussion of practical challenges}. \textit{AStA Advances in
  Statistical Analysis} \textbf{101}(4): 399--438.

\bibitem[{Pohle \textit{et~al.}(2017)Pohle, Langrock, van Beest and
  Schmidt}]{Pohle2017}
Pohle J, Langrock R, van Beest FM, Schmidt NM, 2017. {Selecting the Number of
  States in Hidden Markov Models: Pragmatic Solutions Illustrated Using Animal
  Movement}. \textit{Journal of Agricultural, Biological, and Environmental
  Statistics} \textbf{22}(3): 270--293.

\bibitem[{Quevillon \textit{et~al.}(2015)Quevillon, Hanks, Bansal and
  Hughes}]{Quevillon2015}
Quevillon LE, Hanks EM, Bansal S, Hughes DP, 2015. {Social, spatial, and
  temporal organization in a complex insect society}. \textit{Scientific
  Reports} \textbf{5}: 13393.

\bibitem[{{R Core Team}(2016)}]{RCoreTeam2016}
{R Core Team}, 2016. {R Core Team (2016). R: A language and environment for
  statistical computing.} \textit{R Foundation for Statistical Computing,
  Vienna, Austria. URL http://www.R-project.org/.} .

\bibitem[{Richardson \textit{et~al.}(2017)Richardson, Liechti, Stroeymeyt,
  Bonhoeffer and Keller}]{Richardson2017}
Richardson TO, Liechti JI, Stroeymeyt N, Bonhoeffer S, Keller L, 2017.
  {Short-term activity cycles impede information transmission in ant colonies}.
  \textit{PLoS Computational Biology} \textbf{13}(5): 1--17.

\bibitem[{Schliehe-Diecks \textit{et~al.}(2012)Schliehe-Diecks, Kappeler and
  Langrock}]{Schliehe-Diecks2012}
Schliehe-Diecks S, Kappeler PM, Langrock R, 2012. {On the application of mixed
  hidden Markov models to multiple behavioural time series}. \textit{Interface
  Focus} \textbf{2}(2): 180--189.

\bibitem[{Shaby and Wells(2010)}]{Shaby2010}
Shaby B, Wells M, 2010. {Exploring an adaptive Metropolis algorithm}.
  \textit{Currently under review} : 1--31.

\bibitem[{Tanner and Wong(1987)}]{Tanner1987}
Tanner MA, Wong WH, 1987. {The Calculation of Posterior Distributions by Data
  Augmentation: Rejoinder}. \textit{Journal of the American Statistical
  Association} \textbf{82}(398): 548--550.

\bibitem[{Towner \textit{et~al.}(2016)Towner, Leos-Barajas, Langrock, Schick,
  Smale, Kaschke, Jewell and Papastamatiou}]{Towner2016}
Towner AV, Leos-Barajas V, Langrock R, Schick RS, Smale MJ, Kaschke T, Jewell
  OJ, Papastamatiou YP, 2016. {Sex-specific and individual preferences for
  hunting strategies in white sharks}. \textit{Functional Ecology}
  \textbf{30}(8): 1397--1407.

\bibitem[{van~de Kerk \textit{et~al.}(2015)van~de Kerk, Onorato, Criffield,
  Bolker, Augustine, Mckinley and Oli}]{VandeKerk2015}
van~de Kerk M, Onorato DP, Criffield MA, Bolker BM, Augustine BC, Mckinley SA,
  Oli MK, 2015. {Hidden semi-Markov models reveal multiphasic movement of the
  endangered Florida panther}. \textit{Journal of Animal Ecology}
  \textbf{84}(2): 576--585.

\bibitem[{van Dyk and Meng(2001)}]{VanDyk2001}
van Dyk Da, Meng XL, 2001. {The Art of Data Augmentation}. \textit{Journal of
  Computational and Graphical Statistics} \textbf{10}(1): 1--50.

\bibitem[{Zucchini and MacDonald(2009)}]{Zucchini2009a}
Zucchini W, MacDonald I, 2009. \textit{{Hidden Markov Models for Time Series}},
  volume 110. 278 pp.

\end{thebibliography}

\end{document}